\documentclass[%
 reprint, superscriptaddress,
 amsmath,amssymb,
 aps,
 onecolumn,
]{revtex4-2}

\usepackage{graphicx,dcolumn,bm,hyperref,subfig,color,mathtools}
\usepackage{amsmath, amssymb, gensymb, mathrsfs, bm}
\usepackage{phd_notations}

\begin{document}

\title{Capillary imbibition in a diverging flexible channel}

\author{Mouad Boudina}
\affiliation{Department of Mechanical Engineering, Institute of Applied Mathematics,\\
University of British Columbia, Vancouver, BC, V6T 1Z4, Canada}
\author{Gwynn J. Elfring}%
 \email{gelfring@mech.ubc.ca}
\affiliation{Department of Mechanical Engineering, Institute of Applied Mathematics,\\
University of British Columbia, Vancouver, BC, V6T 1Z4, Canada}

\date{\today}

\begin{abstract}
We study the imbibition of a wetting liquid between flexible sheets that are fixed on both ends. Assuming a narrow gap between the sheets, we solve the lubrication equation coupled with slender body deformation. When the sheets are parallel, we find that the deformation initially speeds up the flow, as shown in previous studies, but only up to the middle of the channel. As the channel contracts, the hydrodynamic resistance increases and ultimately slows down the filling process. Below a threshold stiffness, the channel collapses and imbibition stops. We propose a scaling of the filling duration near this threshold. Next we show that if the sheets are initially tilted with a minimal angle, the channel avoids collapse. The liquid front pulls the diverging sheets and spreads in a nearly parallel portion, which maintains the capillary propulsion and enhances the wicking. Therefore, while it is established that diverging rigid plates imbibe liquids slower than parallel ones, we show that elasticity reverses this principle: diverging flexible sheets imbibe liquids faster than parallel ones. We find an optimal tilt angle that gives the shortest filling time.
\end{abstract}

\maketitle

\section{Introduction}
Transporting a liquid in a macroscopic channel requires an external pump, but channels as thin as hair can move liquids passively without an external effort due to surface tension. Spontaneous capillary pumping can rise sap very high in vegetation with bifurcated conduits \citep{denny2012}, and is the main strategy hummingbirds rely on to extract nectar from flowers \citep{kingsolver1983}. This effortless method is commonly utilized by the microfluidics community to design circuits for fast and simultaneous reagent delivery \citep{juncker2002}, enzyme deposition \citep{tseng2004} or micro-dissected tissue manipulation \citep{astolfi2016}.
The meniscus of a liquid wetting a thin channel results in a pressure jump $\Delta p = \gamma\kappa$ called the Young-Laplace pressure, which drives the motion of the liquid along the channel. The quantity $\gamma$ denotes the surface tension of the liquid, and $\kappa$ denotes the sum of principal curvatures of the meniscus, which is negative for a wetting liquid. During imbibition, the driving pressure balances the viscous resistance, and leads to a square root evolution in time of the meniscus position $\xm \sim t^{1/2}$. The 1/2 exponent was found by \citet{bell1906}, \citet{lucas1918} calculated the prefactor, and \citet{washburn1921} rederived the complete formula.

The motion of liquid is different in channels with a nonuniform cross-section. In general there is a competition between the effect of geometry on the capillary driving force and on the hydraulic resistance. If the walls are undulated, the liquid front speeds up and slows down periodically as it crosses a throat or a bulge \citep{staples2002}. It preserves though, on average, the square root evolution $\xm \sim t^{1/2}$ \citep{lei2021}. In a converging channel, imbibition can be faster than in a straight one, and the filling time is halved for a linearly decreasing cross-section \citep{gorce2016}.  A relevant industrial example is fluids penetrating a converging fracture, which are subject to a hydraulic resistance that increases in the direction of the crack tip \citep{lu2021}. Imbibition in a diverging channel, conversely, is slower than in a straight one. For a linearly diverging gap $h(x) \sim x$, the meniscus advances as $\xm \sim t^{1/3}$, and for a gap increasing as $h(x) \sim x^{n}$, the meniscus advances as $\xm \sim t^{1/(2n+1)}$ \citep{reyssat2008}. In general, the shape of the channel can be optimized, for instance to maximize the final height for a given amount of time during a capillary rise \citep{figliuzzi2013}. The optimal shape is not always converging; it can be a bottleneck having a converging part and a diverging part.

Nonuniform capillaries are common in nature, and the shape variation can lead to optimal hydraulic transport such as in the needles of pine species \citep{zwieniecki2006}. In biological settings, capillaries are often flexible and may deform depending on the flow conditions. Soft tubular channels can buckle due to surface tension while evaporating \citep{hoberg2014}. Shallow channels with a flexible top wall can bulge and expand the flow rate many times \citep{gervais2006}, even if the wall is thick \citep{wang2019}. The deformation of the capillary perturbs the flow properties and can yield erroneous interpretations of the microfluidics measurements. The leading-order increase in the flow rate is a cubic power of the channel gap, and inversely proportional to the Young's modulus of the material \citep{christov2018}.
If the walls of a channel are both elastic, with a fixed base and free tip, they can coalesce during capillary imbibition \citep{aristoff2011}. Collective coalescence is observed as liquid dries off an array of parallel sheets \citep{chiodi2010} or a forest of nanopillars \citep{chakrapani2004, pokroy2009, chandra2009, duan2010, shin2018, ha2022}, forming clusters with special hierarchical patterns. The common feature in this literature is a flexible channel that was initially straight and uniform. Just as capillary flow in nonuniform rigid channels behaves differently than in straight ones, there is reason to believe that starting from a nonuniform flexible channel alters, in a nontrivial manner, the capillary dynamics reported in previous works.

Flows between fixed-free sheets and rods have been the focus of several studies in the elastocapillarity literature. They served as a model for observations of the clustering of wet hair, eyelashes or feathers \citep{bico2004, py2007, duprat2012}, twisting of drying fibers \citep{kovanko2019}, ink uptake of a paint brush \citep{kim2006, wang2014, wang2015a}, water absorption by a paper towel \citep{nasouri2019}, or domino-like liquid transport in moss \citep{ha2021}. Also, fixed-free sheets always guide a droplet toward the tips in a controllable time, whether the liquid is hydrophilic or hydrophobic \citep{bradley2019}.
Soft channels with fixed ends, although less studied, are of research interest and may be of relevance to living systems such as capillary blood vessels, hummingbird tongues \citep{kim2012a}, and insect probosces \citep{krenn2005}. In fact, in paper-based microfluidics, sandwiching a channel between two flexible films accelerates the flow an order of magnitude \citep{jahanshahi-anbuhi2012}. Moreover, a microgroove sealed with a flexible membrane drives the liquid front according to the classical law $\xm \sim t^{1/2}$, yet with a prefactor higher than in rigid rectangular channels \citep{vanhonschoten2007, anoop2015}. One proposed explanation of this improvement is that the membrane deflection shrinks the meniscus, hence increases the capillary pressure. As the membrane keeps deflecting, the channel narrows, increases the hydraulic resistance, and slows down the wicking. But is the capillary propulsion strong enough to overcome the hydraulic resistance? This proposed explanation highlighted only one of the two conflicting effects, therefore needs to be revised.

The present paper studies capillary flows in channels of nonuniform sections and flexible walls at the same time.
We first describe the physics of capillary imbibition between two flexible sheets initially parallel. We determine the threshold stiffness below which the channel collapses, and propose a formula for the filling time. Then, we explain how initially tilting the sheets prevents collapse and enhances the elastocapillary wicking. For a given stiffness, we give the variation of the minimal tilt to avoid collapse, as well as the optimal tilt that yields the fastest imbibition.

\section{Model}
\begin{figure}
\centering
\includegraphics{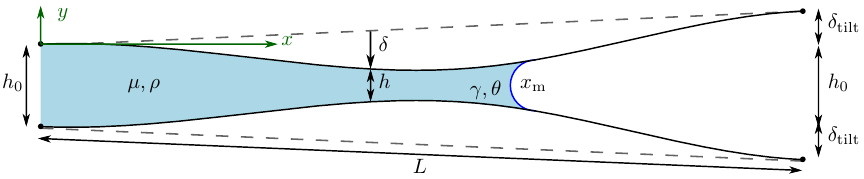}
\caption{Schematics of a liquid imbibition between initially diverging flexible sheets. Dashed lines indicate the state of the sheets before deformation.}
\label{fig:scheme}
\end{figure}

We consider the problem of a Newtonian liquid of dynamic viscosity $\mu$ flowing in a two-dimensional capillary channel made of two flexible sheets of length $L$, as shown in Figure \ref{fig:scheme}. The sheets are clamped on both sides, and separated by a distance $h_{0}$ on the base side and $h_0 + 2\deltatilt$ on the tip side. Both $h_{0}$ and $\deltatilt$ are much smaller than $L$. We define the ratio $\epsilon = h_{0}/L \ll 1$.
The contact angle between the liquid and the sheets is $\theta$, and we assume that the advancing and receding angles are equal. The system is symmetric with respect to the midline of the channel, hence we will focus on the deformation of the upper sheet alone. Let $\delta(x,t)$ denote the deflection of the sheets from the initial configuration at abscissa $x$ and time $t$. The expression of the gap is
\begin{equation}
h(x,t) = h_{0} + 2[\delta(x,t) + \deltatilt x/L].
\label{eq:gap}
\end{equation} 
Note that the $y$-axis points upward, therefore a contraction of the channel means a negative~$\delta$.

\subsection{Structural part}
The liquid exerts on the sheets an internal pressure $p$. Taking the atmospheric pressure as a reference, the linear equation of a slender body deformation is \citep{landau1970}
\begin{equation}
B\parDer{^{4}\delta}{x^{4}} =
\begin{cases}
&p, \quad x \in [0, \xm], \\
&0, \quad x \in [\xm, L],
\end{cases}
\label{eq:p_delta4}
\end{equation}
where $B$ is the bending modulus per unit width and $\xm$ the meniscus position. The boundary conditions for a sheet clamped on both ends are
\begin{equation}
\delta = 0,\quad \parDer{\delta}{x} = 0,\quad \text{ at } x=0,L.
\end{equation}
At the meniscus position $\xm$, the deflection $\delta$ and deflection angle $\partial\delta/\partial x$ are continuous. In absence of external torques, the second derivative $\partial^{2}\delta/\partial x^{2}$ is also continuous. However, the third derivative $\partial^{3}\delta/\partial x^{3}$ is discontinuous owing the downward contact line force $-\gamma \sin(\theta + \partial_{x}\delta)$. Finally, the discontinuity of the fourth derivative $\partial^{4}\delta/\partial x^{4}$ comes from the Young-Laplace pressure drop $2\gamma\cos(\theta + \partial_{x}\delta + \deltatilt/L)/h$. We have in summary
\begin{equation}
\begin{gathered}
\left[ \delta \right]_{\xm^{-}}^{\xm^{+}} = 0,\quad
\left[ \parDer{\delta}{x} \right]_{\xm^{-}}^{\xm^{+}} = 0,\quad
\left[ \parDer{^{2}\delta}{x^{2}} \right]_{\xm^{-}}^{\xm^{+}} = 0,\\
\left[ \parDer{^{3}\delta}{x^{3}} \right]_{\xm^{-}}^{\xm^{+}} = -\frac{\gamma}{B}\sin(\theta + \partial_{x}\delta|_{\xm} + \deltatilt/L),\\
\left[ \parDer{^{4}\delta}{x^{4}} \right]_{\xm^{-}}^{\xm^{+}} = \frac{2\gamma\cos(\theta + \partial_{x}\delta|_{\xm} + \deltatilt/L)}{Bh|_{\xm}}
\end{gathered} 
\end{equation}
where the superscripts $\pm$ in $\xm$ denote the limit value from the right and left sides of the meniscus. It should be noted that the contact line force generally pinches the soft substrate on which it lies and creates a localized bump \citep{marchand2012}. Nonetheless, since the sheets are very thin, the size of the bump is negligible compared to the size of the structure deformation \citep{andreotti2020}.

In the dry part of the channel, $x \in [\xm,L]$, the sheets are free of load, $\partial^{4}\delta/\partial x^{4} = 0$. Integrating and evaluating the clamped boundary conditions at $x = L$, we obtain
\begin{align}
\delta(x,t) &= \delta|_{\xm} \left[ 3 \left( \frac{L-x}{L-\xm} \right)^{2} - 2\left( \frac{L-x}{L-\xm} \right)^{3} \right]
+ (L-\xm)\partial_{x}\delta|_{\xm} \left[ \left( \frac{L-x}{L-\xm} \right)^{2} - \left( \frac{L-x}{L-\xm} \right)^{3} \right].
\end{align}
From this expression we deduce the two remaining boundary conditions in the wet part
\begin{align}
&\parDer{^{2}\delta}{x^{2}}\biggr|_{\xm^{-}} = - \frac{6\delta|_{\xm}}{(L-\xm)^{2}} - \frac{4\partial_{x}\delta|_{\xm}}{L-\xm} \label{eq:bc2} \\
&\parDer{^{3}\delta}{x^{3}}\biggr|_{\xm^{-}} = \frac{12\delta|_{\xm}}{(L-\xm)^{3}} + \frac{6\partial_{x}\delta|_{\xm}}{(L-\xm)^{2}}
+ \frac{\gamma}{B}\sin(\theta + \partial_{x}\delta|_{\xm}).
\label{eq:bc3}
\end{align}

\subsection{Fluid part}
In the lubrication approximation, the flow rate per unit width is $q = -(h^{3}/12\mu)\partial p/\partial x$, which, combined with \eqref{eq:p_delta4}, equals
\begin{equation}
q = -\frac{Bh^{3}}{12\mu} \parDer{^{5}\delta}{x^{5}}.
\label{eq:flow_rate}
\end{equation}
Mass conservation requires that
\begin{equation}
\parDer{h}{t} + \parDer{q}{x} = 0.
\label{eq:mass_conservation}
\end{equation}

Upon substituting \eqref{eq:flow_rate} in \eqref{eq:mass_conservation}, and given the expression of the gap \eqref{eq:gap}, we get
\begin{equation}
\parDer{\delta}{t} = \frac{B}{24\mu} \parDer{}{x} \left[ h^{3} \parDer{^{5}\delta}{x^{5}} \right]
= \frac{B}{24\mu} \left( h^{3} \parDer{^{6}\delta}{x^{6}} + 3h^{2} \parDer{h}{x} \parDer{^{5}\delta}{x^{5}} \right).
\label{eq:hdot}
\end{equation}

As for the advancing front, we have
\begin{equation}
\totDer{\xm}{t} = \frac{q}{h}\biggr|_{\xm} = -\frac{Bh^{2}}{12\mu} \parDer{^{5}\delta}{x^{5}} \biggr|_{\xm}.
\label{eq:xmdot}
\end{equation}

The initial conditions are
\begin{equation}
\delta(x,0) = 0,\quad
\xm(0) = x_{\mathrm{m},0}.
\end{equation}

The initial position $x_{\mathrm{m},0}$ should be chosen such that imbibition happens after the inertial regime.
For this reason we consider the instant $t \sim \rho L^{2}/\mu$ at which the expression of the meniscus position in viscous regime \citep{zhmud2000}, $\sqrt{h_{0}\gamma t\cos\theta/3\mu}$, overlaps with the one in the inertial regime \citep{fries2008}.
This means that the initial position should be greater than $L\sqrt{\rho h_{0}\gamma\cos\theta/3\mu^{2}}$. We choose a starting position of $x_{\mathrm{m},0}/L = 10^{-2}$.

\subsection{Non-dimensionalization}
We non-dimensionalize the longitudinal distances $x$ and $\xm$ by $L$, and the cross-sectional distances $h$ and $\delta$ by $h_{0}$. We scale the time $t$ by the duration it takes to fill parallel rigid sheets with gap $h_0$, $\tau_0 = 3\mu L^{2}/h_{0}\gamma\cos\theta$. Denoting the dimensionless variables with a tilde,
the channel gap is rewritten as
\begin{equation}
\htld(\xtld,\ttld) = 1 + 2[\deltatld(\xtld,\ttld) + \deltatilttld\xtld],
\end{equation}
and the equations become
\begin{align}
&\parDer{\deltatld}{\ttld}
= \frac{\calNec}{8} \left( \htld^{3} \parDer{^{6}\deltatld}{\xtld^{6}} + 3\htld^{2} \parDer{\htld}{\xtld} \parDer{^{5}\deltatld}{\xtld^{5}} \right)
\label{eq:deltadot_dless},\\
&\totDer{\xmtld}{\ttld} = -\frac{\calNec}{4} \htld^{2}\parDer{^{5}\deltatld}{\xtld^{5}}\biggr|_{\xmtld},
\label{eq:xmdot_dless}
\end{align}
The dimensionless group that naturally emerges is the elastocapillary number \citep{mastrangelo1993}
\begin{align}
\calNec = \frac{Bh_{0}^{2}}{\gamma \cos\theta L^4}
\end{align}
This number may also be viewed as the ratio of length scales $\calNec = (\lec/L)^4$ where $\lec = (Bh_{0}^{2}/\gamma \cos\theta)^{1/4}$ is the elastocapillary length which balances the bending energy with the surface energy $B\lec(h_{0}/\lec^{2})^{2} \sim \gamma\lec$.
When $\calNec \gg 1$ the channel is essentially rigid, and decreasing values mean relatively more flexible channel walls. When $\calNec \ll 1$ one expects deformation over a length scale much smaller than the length of the channel as $\lec \ll L$.
At some critical value of the elastocapillary number, $\calNec^{*}$, the sheets become flexible enough and the channel walls touch.

\subsection{Numerical solution}
We solve the equation in the wet domain using the finite difference method. The discretisation scheme is given in the Appendix. We transform the equations \eqref{eq:xmdot_dless} using the normalised variable $\xtld/\xmtld$ to avoid integration over an elongating domain $[0, \xmtld(\ttld)]$, then numerically integrate in the unit domain~$[0, 1]$ that we discretise into $\Nx$ cells. We use an implicit Euler scheme with respect to time, and a second-order central difference scheme with respect to space. We introduce three ghost points on each end to account for the boundary conditions. The two equations are rearranged into a nonlinear matrix equation, which we linearise and solve using Picard iterations with a relaxation parameter of 0.7 \citep{langtangen2017}. The procedure terminates if the difference between two successive iterations is less than $10^{-5}$. We stop the simulation when the meniscus reaches the tip, $\xmtld = 1$, or when the sheets touch, $\min_{\xtld} \htld < 10^{-3}$. The results presented in the following are obtained for $\Nx = 100$ and a rescaled time step of $(\calNec/4)\Delta \ttld = 10^{-5}$. Throughout this work, we take a gap-to-length ratio of $\epsilon = 0.05$ and a contact angle of $\theta = 0$.

\section{Results (with Discussion)}
\subsection{Parallel sheets}
\subsubsection{Deformation profiles}
\begin{figure}
\centering
\includegraphics{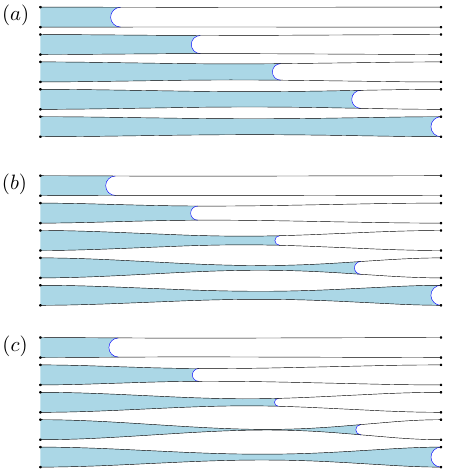}
\caption{Liquid imbibition between sheets of ($a$) $\calNec = 2.56 \times 10^{-2}$, $(b)$ 0.81 $\times 10^{-2}$, and ($c$) 0.65 $\times 10^{-2}$, right above the collapse threshold $\calNec^{*} \approx 0.64 \times 10^{-2}$. The softest sheets are the fastest to fill half the channel, but the slowest to fill the rest. The gap-to-length ratio is $\epsilon = 0.05$. The snapshots are taken at positions $\xmtld$ = 0.2, 0.4, 0.6, 0.8, 1, corresponding to times ($a$) $\ttld =$ 0.04, 0.16, 0.33, 0.67, 1.29, ($b$) 0.04, 0.14, 0.27, 1.36, 5.91, and ($c$) 0.05, 0.13, 2.58, 10.43, 86.27. The axes are to scale.}
\label{fig:snaps_straight}
\end{figure}

\begin{figure}
\centering
\includegraphics{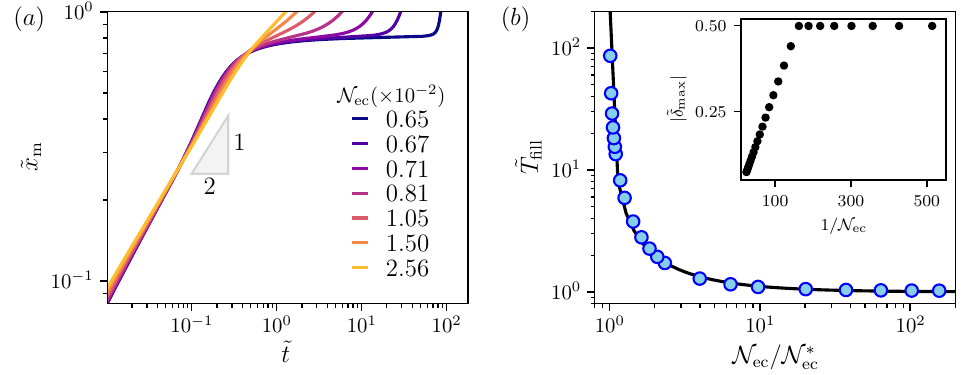}
\caption{($a$) Meniscus position versus time for parallel sheets of different elastocapillary numbers. ($b$) Filling time versus the normalized elastocapillary number. The solid line is the relation \eqref{eq:Tfill_sim_hmin}. Inset: maximum deflection $\abs{\deltamax}$ versus $1/\calNec$.}
\label{fig:xm_vs_time}
\end{figure}

We consider the case of initially parallel sheets $\deltatilttld = 0$. Figure \ref{fig:snaps_straight} shows snapshots of imbibition for three elastocapillary numbers. In the case $\calNec = 2.56 \times 10^{-2}$ ($\lec/L = 0.4$), the channel almost preserves its parallel state, except an inappreciable contraction in the middle. As shown in Figure \ref{fig:xm_vs_time}($a$), the position of the meniscus closely follows a power law of exponent 1/2, like in traditional imbibition, until the midway point. After that, however, the flow slows down and fills the channel at $\Tfilltld = 1.29$, longer than between rigid plates.

Imbibition between softer sheets $\calNec = 0.81 \times 10^{-2}$ is represented in Figure \ref{fig:snaps_straight}($b$). Prior to halfway, the sheets deflect and form a converging portion, which allows the liquid to spread faster than in rigid sheets. A converging profile speeds up the flow because the sheets shrink the meniscus, hence increase the driving Young-Laplace pressure.
Yet, this enhancement does not last after crossing halfway, since the sheets form a narrow neck,
increasing the hydraulic resistance and retarding the flow in the remaining half. Only towards the end of the channel does the meniscus speed up, owing to the neck relaxation and hydraulic resistance reduction. The filling time is $\Tfilltld \approx 6$, i.e. six times longer than in the rigid case.

\subsubsection{Filling time}
Figure \ref{fig:xm_vs_time}($b$) shows the evolution of the channel filling time $\Tfilltld$ with the elastocapillary number. We recover the value of traditional imbibition between rigid plates $\Tfillinftytld = 1$ for high enough $\calNec$. As the elastocapillary number decreases, the filling time increases and sharply diverges at a critical value $\calNec^{*}$. In this state, the channel takes an infinite time to fill because the gap constricts $\hmintld \rightarrow 0$, yielding an infinite hydraulic resistance, hence a zero flow rate. The critical elastocapillary number $\calNec^{*}$ is the threshold below which the channel collapses. Numerically we find $\calNec^{*} \approx 0.64 \times 10^{-2}$. Figure \ref{fig:snaps_straight}$(c)$ depicts the imbibition between sheets of $\calNec = 0.65 \times 10^{-2}$, just above the collapse condition. From Figure \ref{fig:xm_vs_time}($a$), the constricted state is what takes the most of the filling time.

A doubly-fixed beam subject to a force $f$ per unit length and width has, at the midpoint, a maximal deflection of $fL^{4}/384B$. If we replace the distributed force $f$ by $2\gamma/h_{0}$, it becomes $\gamma L^{4}/192Bh_{0}$, which reads in dimensionless form $1/192\calNec$. In our case, although the load distribution applied by the liquid varies in time, we still expect that $\abs{\deltamaxtld} \propto 1/\calNec$.

Assuming the formal relationship
\begin{equation}
\deltamaxtld = -\frac{1}{2} \frac{\alpha}{\calNec},
\end{equation}
and fitting the constant of proportionality $\alpha$ to the simulation data (inset of Figure~\ref{fig:xm_vs_time}$b$), we find that $\alpha = 0.62 \times 10^{-2}$. The threshold elastocapillary number (when $\hmintld \rightarrow 0, \deltamaxtld = -1/2$) is then $\calNec^{*} = \alpha$, and the minimum gap can be written in the form
\begin{equation}
\hmintld = 1 + 2\deltamaxtld = 1 - \frac{\calNec^{*}}{\calNec}.
\label{eq:hmintld}
\end{equation} 

The steep increase of $\Tfilltld \rightarrow \infty$ as $\hmintld \rightarrow 0$ (when $\calNec \rightarrow \calNec^{*}$) suggests that the filling time depends on an inverse power of the minimum gap. Near this singularity, the filling time is determined by the duration the system stays in the constricted state. If we simply take the filling time and naively replace the gap thickness $h_{0}$ with the minimum thickness $\hmin$ found in \eqref{eq:hmintld}, we obtain the following formula for the dimensionless filling time
\begin{equation}
\Tfilltld = \frac{1}{\hmintld} = \frac{1}{1 - \calNec^{*}/\calNec},
\label{eq:Tfill_sim_hmin}
\end{equation}
which matches the numerical results in Figure~\ref{fig:xm_vs_time}($b$) exceptionally well over the entire span of the data.

\subsection{Tilted sheets}
\subsubsection{Deformation profiles}
\begin{figure}
\centering
\includegraphics{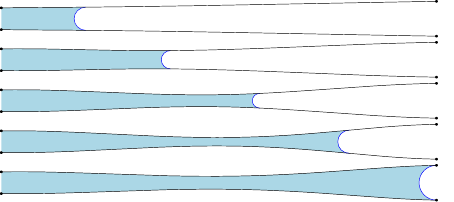}
\caption{Liquid imbibition between sheets initially tilted with $\deltatilttld = 0.3$, of $\calNec = 0.53 \times 10^{-2}$ below the threshold $\calNec^{*}$. The gap-to-length ratio is $\epsilon = 0.05$. The snapshots are taken at positions $\xmtld$ = 0.2, 0.4, 0.6, 0.8, 1, corresponding to times $\ttld =$ 0.05, 0.14, 0.28, 1.56, 4.91. The axes are to scale.}
\label{fig:snaps_tilted}
\end{figure}

We found that the straight soft sheets having an elastocapillary number below $\calNec^{*}$ contract quickly and collapse. But if the same soft sheets are initially tilted, the channel avoids collapse, as shown in Figure \ref{fig:snaps_tilted}. The meniscus advances in a weakly converging portion until filling half of the channel, then crosses to a diverging portion and fills the rest slowly.

\begin{figure}
\centering
\includegraphics{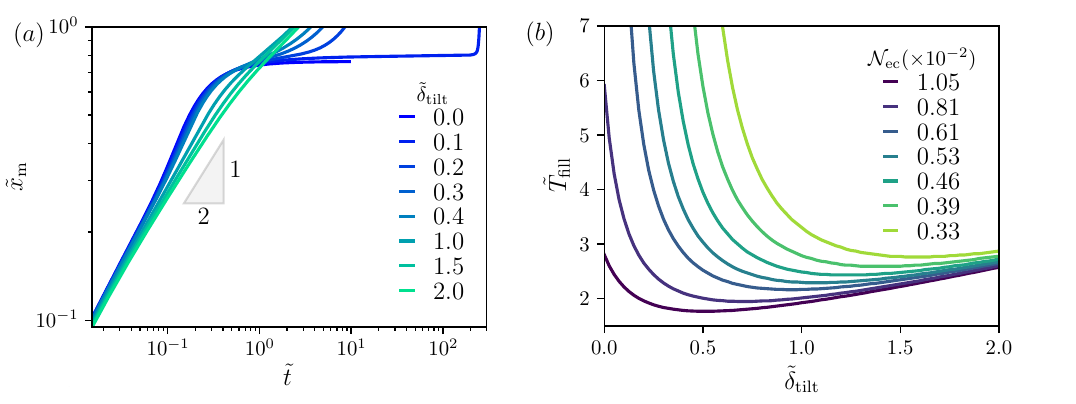}
\caption{($a$) Meniscus position in time for diverging sheets of different tilts, with $\calNec = 0.53 \times 10^{-2}$. ($b$) Filling time versus tilt for different elastocapillary numbers. The curves of $\calNec = 1.05 \times 10^{-2}$ and 0.81 $\times 10^{-2}$ have a finite value at $\deltatilttld = 0$, because they are above the threshold elastocapillary number $\calNec^{*}$. Below, the filling time diverges at a nonzero critical tilt $\deltatilttld^{*} > 0$.}
\label{fig:xm_vs_time_tilt}
\end{figure}

The meniscus position in time for several tilts is represented in Figure \ref{fig:xm_vs_time_tilt}($a$). For small tilts, the evolution of $\xmtld$ shares features similar to the parallel case, namely a speed-up before halfway and a slow-down after, with a short increase in speed at the end due to the neck relaxation. For high tilts, the meniscus does not speed up then slow down, but instead maintains a uniform speed throughout the imbibition process.

\subsubsection{Filling time, critical and optimal tilts}
\begin{figure}
\centering
\includegraphics{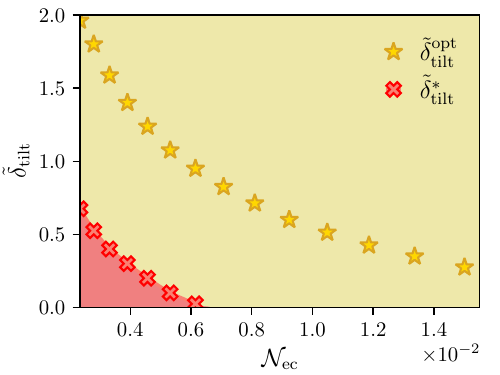}
\caption{Phase diagram ($\deltatilttld, \calNec$), with the variation of the critical tilt $\deltatilttld^{*}$ (crosses) and optimal tilt $\deltatiltopttld$ (stars). The channel collapses in the red region below the critical tilt, and stay open in the golden region above. The critical curve can be also read as the threshold elastocapillary number $\calNec^{*}$ decreasing for high tilts.}
\label{fig:tilt_star_opt_vs_lec}
\end{figure}

Figure \ref{fig:xm_vs_time_tilt}($b$) shows the filling time versus the tilt for several elastocapillary numbers. We notice the existence of two tilts of interest.
The first is the critical tilt $\deltatilttld^{*}$, which forms the separatrix between collapsed and uncollapsed states, i.e. between infinite and finite filling times.
The value of $\deltatilttld^{*}$ depends on the stiffness of the sheets as shown in Figure \ref{fig:tilt_star_opt_vs_lec}. For sheets having $\calNec > \calNec^{*}$, it equals $\deltatilttld^{*} = 0$ because the channel stays open anyways if initially straight. As the elastocapillary number decreases below $\calNec^{*}$, the critical tilt increases, meaning that the sheets have to be more diverging in order to ensure a safety margin and avoid collapse.

\begin{figure}
\centering
\includegraphics{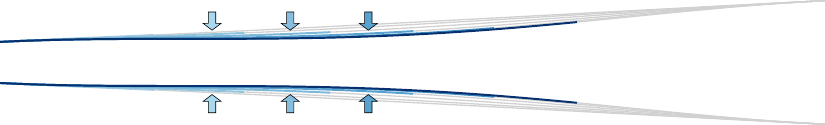}
\caption{Deformation profiles of the sheets in the case $\calNec = 0.53 \times 10^{-2}$ at the optimal tilt angle $\deltatiltopttld \approx 1$. The blue and gray lines indicate the wet and dry portions. The snapshots are taken at positions $\xmtld = 0.2$ (light blue), 0.3, 0.4, 0.5, 0.6 and 0.7 (dark blue), and the axes are to scale.}
\label{fig:deformation_optimal_tilt}
\end{figure}

The second tilt of interest is the optimal tilt $\deltatiltopttld$ that gives the fastest filling of the channel (Figure~\ref{fig:tilt_star_opt_vs_lec}). For illustration, Figure~\ref{fig:deformation_optimal_tilt} shows the deformation profiles of the case $\calNec = 0.53 \times 10^{-2}$ at the optimal tilt $\deltatiltopttld \approx 1$. The liquid front pulls the sheets and spreads between nearly parallel portions, until filling more than half the channel. This maintains the same meniscus size, hence the same capillary driving pressure throughout the wicking. After crossing halfway, the liquid front spreads in a slightly diverging portion, but without slowing down given the relaxation of the neck and concomitant reduction of the hydraulic resistance. This configuration is reminiscent of the optimised capillary shape by \citet{figliuzzi2013}, which was not entirely converging but had a long converging part and a short diverging part near the ends.

\section{Conclusion}
This paper investigated the capillary flow between two elastic sheets fixed at both ends. We first studied the capillary flow between sheets that are initially parallel, and then explored the elastocapillary interaction when the sheets are tilted outwards. In the parallel case, we found that flexibility speeds up the capillary flow, as shown by \citet{vanhonschoten2007} and \citet{anoop2015}, yet only before reaching halfway. As the meniscus crosses halfway, the sheets contract and increase the hydraulic resistance, subsequently decelerating the flow. If the stiffness of the sheets is below a minimal threshold, the channel collapses and blocks imbibition. If, however, the sheets are initially tilted with a sufficient angle, the channel avoids collapse. Tilting the sheets hence lowers the threshold elastocapillary number.

The total filling time in a straight flexible channel is longer than in a rigid one, and diverges as the elastocapillary number goes down to a critical value. Our results show that the filling time scales as the inverse of the minimum gap in the channel, just as the filling time between rigid walls scales as the inverse of the channel gap.

When the sheets are tilted, the flow speeds up and the filling time decreases. We found an optimal tilt angle for a given stiffness corresponding to the shortest filling time. It is known that liquid imbibition in diverging plates is slower than in parallel ones \citep{reyssat2008}. When the plates are flexible, however, we showed that this principle reverses: liquid imbibition can be faster in diverging sheets than in parallel ones.

We focused on flexible channels with linearly diverging cross-sections, with the goal of developing insight into how the simplest geometrical nonuniformity affects capillary imbibition and compares with the traditional problem of parallel sheets. Other nonuniformities that could be examined are sheets with intrinsic curvatures, or made of materials with heterogeneous porosity, elasticity or wettability gradients.

Standard microfluidic channels, which are fabricated by 3D-printing or laser cutting, have thick and relatively nonflexible walls, as well as straight and uniform cross-sections. Starting from these constraints, a collection of capillaric elements were invented to control and manipulate the microfluidic spreading of fluids like flow resistors, capillary pumps and valves \citep{olanrewaju2018}. Outside of these constraints, a flexible channel with tunable tilt angle might be introduced as a capillaric element with a dual functionality. In the parallel configuration, it can serve as a delay valve that slows down the flow, or a stop valve that closes the channel if the elastocapillary number is lower than the collapse threshold, for instance when the walls are loose or long enough. This single-device configuration would be more convenient than the conventional multi-level delay valves that require space \citep{zimmermann2008}. Then, in the diverging configuration, tuning the tilt angle to the optimal value may act as a speed amplifier. It would be relevant to seek new methods of fabricating channels with soft slender walls, and embed them in circuits along with other capillaric elements. Some biological observations, such as the successive suction pumps in insect probosces \citep{krenn2005} or the domino-like liquid transport in moss \citep{ha2021}, can inspire the study of circuits with flexible diverging walls in series.

The theory presented here is based on the assumption of infinitely wide sheets, neglecting the effect of the liquid free surface on the edges. Nevertheless, in open microchannels, which are easier to fabricate and practical in biomedical applications \citep{oliveira2019}, the free-surface is a relevant factor that may affect the flow \citep{kolliopoulos2021}. It would be fruitful to extend the present work to the case of open fluidics systems, such as a suspended microchannel \citep{casavant2013, berthier2015} with flexible walls, which would pave the way for potential new lines of research.
\\\indent
Using open microchannels also means an exposure of the free surface to the ambient air and a subsequent evaporation, which was shown to hinder the flow during imbibition \citep{lade2018} and capillary rise \citep{kim2022a}. Elastocapillary deformation of the channel would alter the geometry of the free surface, hence the exposed area, which may lead to changes in evaporation that deserve to be addressed.

The flow enhancement in diverging ducts reminds the diverging profile of food canals in
bees \citep{borrell2006}, butterflies \citep{krenn2000, krenn2002} and hummingbirds \citep{kim2012}.
These capillaries transport biofluids which carry polymers, vesicles and active particles. Microfluidic devices for medical and chemical applications too deliver reagents and chemicals that are inherently different from Newtonian fluids. An interesting extension of the present study would be to look at the elastocapillary imbibition of fluids displaying non-Newtonian rheology.

We considered sheets that are smooth, which is an ideal configuration compared to the textured walls in real capillaries found in nature. This feature is beneficial for microscale mass transport. In fact, microchannels with herringbone patterned grooves capture better magnetised microbeads \citep{lund-olesen2007}, and generate counter-rotating vortices that mix efficiently the flow \citep{stroock2002, stroock2002a}. Additionally, if a tube has an inner surface mimicking the texture of the tropical pitcher plant, it ascends water to twice the final height of a normal tube, with a speed eighteen times faster \citep{li2019a}. We believe that combining flexibility with texture could bring exciting improvements worth exploring in future projects.

\acknowledgements
The authors gratefully acknowledge funding from the Natural Sciences and Engineering Research Council of Canada (RGPIN-2020-04850). 

\bibliography{phd_biblio}

\section{Appendix: Discretization scheme}
We consider a backward Euler scheme with respect to time, and a centered scheme with respect to space. At the time step $n$, the deformation equation~\eqref{eq:deltadot_dless} is, for $k = 0, \dots, \Nx$,
\begin{align*}
&\frac{2\Delta x^{6}}{\Delta t}
(\delta^{n}_{k} - \delta^{n-1}_{k})
= \frac{\calNec}{4} [ \left.
h^{3}(\delta^{n}_{k}) (\delta^{n}_{k+3} - 6\delta^{n}_{k+2}+15\delta^{n}_{k+1}-20\delta^{n}_{k}+15\delta^{n}_{k-1}-6\delta^{n}_{k-2}-\delta^{n}_{k-3}) \right.\\
&\left.+ 6h^{2}(\delta^{n}_{k}) \left( \frac{\delta^{n}_{k+1} - \delta^{n}_{k-1}}{2\Delta x} + \delta'_{0}(x_{k}) \right)
(0.5\delta^{n}_{k+3} - 2\delta^{n}_{k+2} + 2.5\delta^{n}_{k+1} - 2.5\delta^{n}_{k-1} + 2\delta^{n}_{k-2} - 0.5\delta^{n}_{k-3}) ] \right.,
\end{align*} 
and the meniscus equation~\eqref{eq:xmdot_dless} is
\begin{equation*}
\frac{\Delta x^{5}}{\Delta t} (\xm^{n} - \xm^{n-1}) = - \frac{\calNec}{4} h^{2}(\delta^{n}_{\Nx})
[0.5\delta^{n}_{\Nx+3} - 2\delta^{n}_{\Nx+2} + 2.5\delta^{n}_{\Nx+1} - 2.5\delta^{n}_{\Nx-1} + 2\delta^{n}_{\Nx-2} - 0.5\delta^{n}_{\Nx-3}],
\end{equation*}
where the tildes $\tilde{(.)}$ are dropped off for convenience.

During integration the time is rescaled to $(\calNec/4)t$. We introduce three ghost points that account for the boundary conditions. From these equations we build a nonlinear matrix system
\begin{equation}
\calA(\bV^{n})\bV^{n} = \bb(\bV^{n}, \bV^{n-1}),
\label{eq:linear_matrix_equation}
\end{equation} 
with $\bV = (\delta_{-3}, \dots, \delta_{\Nx+3}, \xm)$, having $\Nx + 8$ components.

We solve the matrix equation \eqref{eq:linear_matrix_equation} using Picard iterations. We start from an initial guess $\bV^{*}$, usually taken as the solution at the previous time step $\bV^{n-1}$, and linearise the left hand side. We solve the equation $\calA(\bV^{*})\bV = \bb(\bV, \bV^{n-1})$, and update the next iteration using a relaxation parameter of $\alpha = 0.7$, i.e. $\bV \leftarrow \alpha\bV + (1-\alpha)\bV^{*}$. If the absolute error between two successive iterations is $\abs{\bV - \bV^{*}} < 10^{-5}$, we stop the iterations and the solution at the current step is $\bV^{n} \leftarrow \bV$.

\end{document}